REVIEW                                                                                                    Open Access

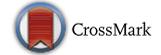

# A new view on the solar wind interaction with the Moon

Anil Bhardwaj[1*], M B Dhanya[1], Abhinaw Alok[1], Stas Barabash[2], Martin Wieser[2], Yoshifumi Futaana[2], Peter Wurz[3], Audrey Vorburger[4], Mats Holmström[2], Charles Lue[2], Yuki Harada[5] and Kazushi Asamura[6]

**Abstract**

Characterised by a surface bound exosphere and localised crustal magnetic fields, the Moon was considered as a passive object when solar wind interacts with it. However, the neutral particle and plasma measurements around the Moon by recent dedicated lunar missions, such as Chandrayaan-1, Kaguya, Chang'E-1, LRO, and ARTEMIS, as well as IBEX have revealed a variety of phenomena around the Moon which results from the interaction with solar wind, such as backscattering of solar wind protons as energetic neutral atoms (ENA) from lunar surface, sputtering of atoms from the lunar surface, formation of a "mini-magnetosphere" around lunar magnetic anomaly regions, as well as several plasma populations around the Moon, including solar wind protons scattered from the lunar surface, from the magnetic anomalies, pick-up ions, protons in lunar wake and more. This paper provides a review of these recent findings and presents the interaction of solar wind with the Moon in a new perspective.

## Introduction

The Moon is a regolith covered planetary object characterised by a surface-bound exosphere (Killen and Ip 1999; Stern 1999; Sridharan et al. 2010) and an absence of a global magnetic field. However, the Moon possesses regions of stronger crustal magnetic fields, known as 'magnetic anomaly' regions (Coleman et al. 1972; Hood and Schubert 1980; Lin et al. 1998; Hood et al. 2001; Halekas et al. 2001; Richmond et al. 2003; Tsunakawa et al. 2015). The Moon was thought to be a passive absorber of the plasma when the supersonic solar wind is incident on its surface (Feldman et al. 2000; Crider and Vondrak 2002). However, observations from the recent missions, such as Chandrayaan-1, Kaguya, Chang'E-1, ARTEMIS, and IBEX have shown that the interaction of solar wind with the Moon is quite dynamic with a variety of processes involved, such as sputtering of lunar surface, scattering of 0.1–1 % solar wind protons with the same charge state and ≤28 % as hydrogen energetic neutral atoms (ENAs), existence of "mini-magnetosphere" at the lunar surface, reflection of solar wind protons from lunar magnetic anomalies, protons in near-lunar wake by different entry mechanisms and so on.

This paper provides an overview of the recent findings in the area of solar wind interaction with the Moon and presents a new perspective on this subject. It also tries to correlate the lunar observations with those conducted in laboratories.

## Review

### Neutral particle environment around the Moon

The observations by the CENA sensor of the SARA experiment onboard Chandrayaan-1 (Bhardwaj et al. 2005, 2010; Barabash et al. 2009) and the IBEX-Lo and IBEX-Hi of IBEX (McComas et al. 2009) have provided new insights into the energetic neutral atom (ENA) environment around the Moon.

Energetic neutral hydrogen atoms (hydrogen ENAs) generated by the interaction of solar wind protons with lunar regolith have been observed by the ENA sensors onboard Chandrayaan-1 (Wieser et al. 2009) and IBEX

*Correspondence: anil_bhardwaj@vssc.gov.in
[1] Space Physics Laboratory, Vikram Sarabhai Space Centre, Trivandrum 695022, India
Full list of author information is available at the end of the article

(McComas et al. 2009b). The estimated backscatter fraction (albedo) of hydrogen ENAs from Chandrayaan-1 and IBEX observations are about 8–28 %. These values were much higher than those expected (Crider and Vondrak 2002). The ENA intensity displayed variations with the solar zenith angle (Fig. 1). The derived angular scattering function for the ENAs using CENA observations (Schaufelberger et al. 2011) suggests more scattering in the sunward direction (backscattering) than the forward scattering as the solar zenith angle increases, which is contrary to the expectations based on laboratory studies (Niehus et al. 1993). Using this scattering function, the global mean for hydrogen ENA albedo from CENA/SARA is 16 ± 5 % (Fig. 2) (Vorburger et al. 2013), while the value estimated from IBEX observations is 11 ± 6 % (Saul et al. 2013).

Futaana et al. (2012) found that the energy spectrum of the backscattered hydrogen ENAs is Maxwellian (Fig. 3) and that the ENA characteristic energies are correlated with the solar wind velocity (rather than solar wind energy). This implied that the ENA emission is a momentum-driven process and the impacting solar wind protons experience multiple collisions within the top layer of the regolith before getting released into the space as neutral hydrogen. The backscatter fraction was computed by integrating the empirical energy spectrum, and its dependence on various solar wind plasma parameters, such as density, velocity, temperature, interplanetary magnetic field, and the fraction of alpha particles have also been investigated (Futaana et al. 2012). It is found that the back-scatter fraction is not controlled by any of the solar wind parameters. However, Funsten et al. (2013) have reported the dependence of the ENA reflection coefficient on the incident solar wind speed from IBEX observations and the ENA reflection ratio ranges from ∼0.2 for slow solar wind to ∼0.08 for fast solar wind. The average energy per incident solar-wind ion reflected to space is found to be ∼30 eV for slow solar wind and ∼45 eV for fast solar wind. Also, from IBEX observations, Allegrini et al. (2013) found the energy spectra to be a power law, and the spectral shape of the ENA emission was represented by a linearly decreasing intensity with increasing energy for energy >250 eV (Funsten et al. 2013).

In the similar way as the interaction of solar wind with Earth's magnetic field results in the formation of magnetosphere which effectively shields the planet from the solar wind, the possibility that magnetic anomaly regions on Moon can form small-scale magnetospheres ("mini-magnetospheres") has been suggested from Lunar Prospector observations (Lin et al. 1998). The discovery of "mini-magnetosphere" on Moon was made possible by means of ENA observations from Chandrayaan-1 over the Crisium antipode magnetic anomaly region (Wieser et al. 2010). The spatial extent of the mini-magnetosphere was found to be ∼360 km across at the lunar surface. The ENA intensity was found to reduce by about 50 % within the area of the mini-magnetosphere. This region of reduced ENA intensity was surrounded by a region of enhanced ENA intensity of width ∼300 km which is due

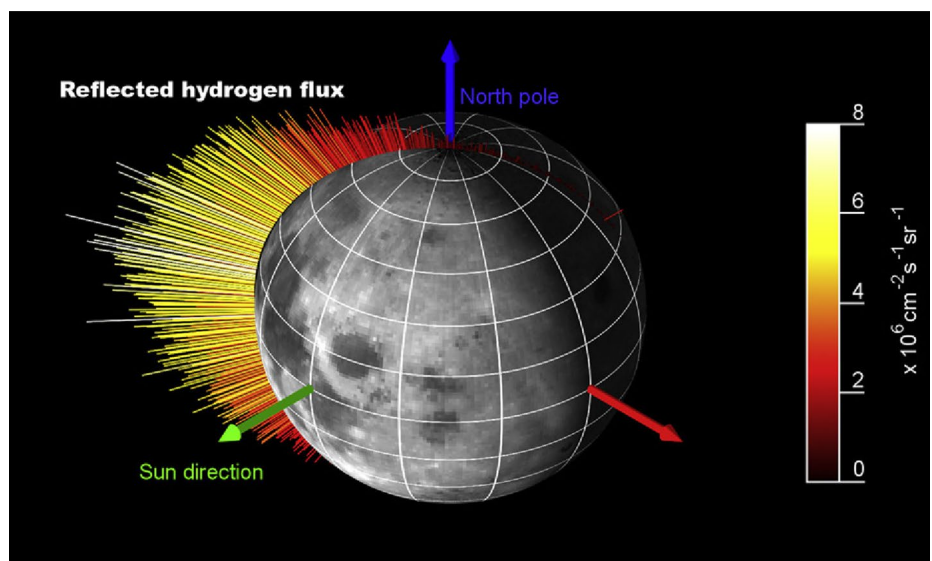

**Fig. 1** The ENA intensity observed by SARA on Chandrayaan-1 on 6 February 2009. The *colour* as well as the length of the *lines* indicate the strength of ENA signal. It can be seen that the ENA intensity varies with the solar zenith angle. The weak signals on night-side is due to instrument background. The lunar surface map is from Clementine image data [from Wieser et al. (2009), Erratum in 2011)]

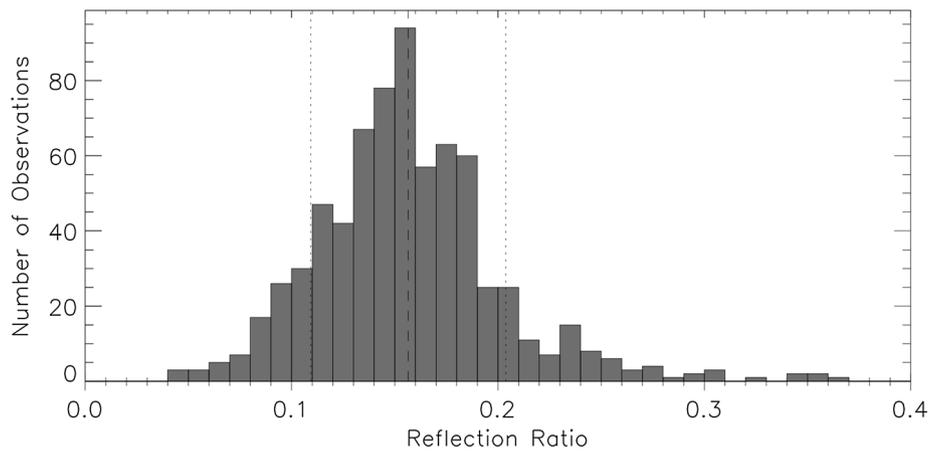

**Fig. 2** The *histogram* showing the reflection ratios for ENA computed from the SARA/CENA observations on Chandrayaan-1. The *vertical dashed line* shows the mean value of 0.16 and the *vertical dotted lines* show the standard deviation of 0.05 [from Vorbuger et al. (2013)]

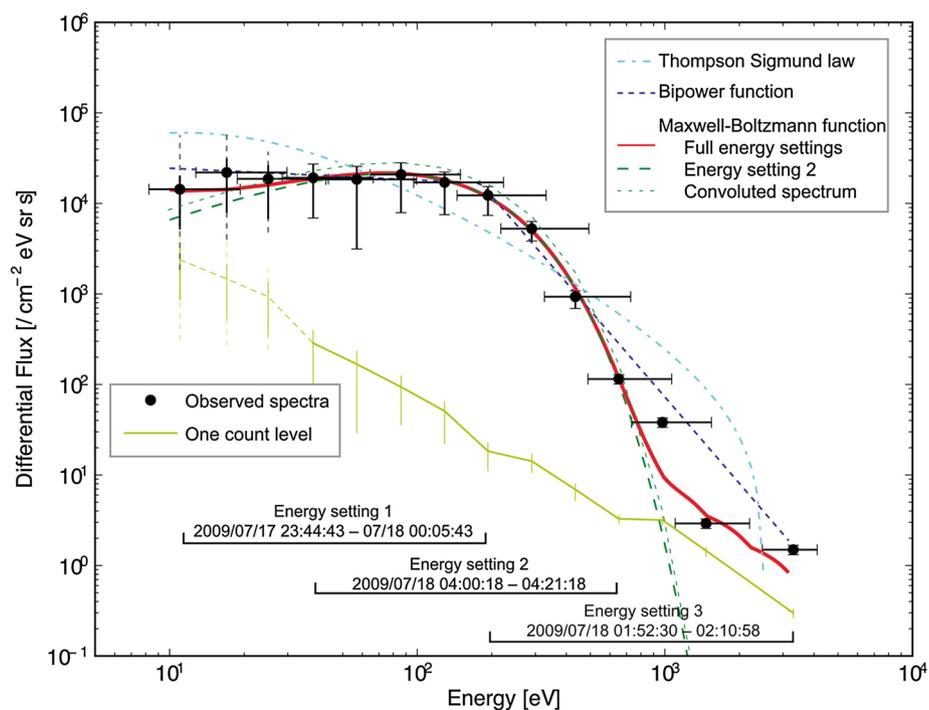

**Fig. 3** The energy spectrum of backscattered ENAs observed by SARA/CENA along with the fitted curves. Data are averaged over 3 orbits during 17–18 July 2009. CENA employed different energy settings and the data for the energy settings 1–3 are indicated. The fits for the Thompson–Sigmund law (*light blue line*), the bi-power law (*blue line*), and the Maxwell–Boltzmann distribution (*red line*) are shown. The *green dashed* and the *green dotted* curves show the fits for the Maxwell–Boltzmann distribution only for the data corresponding to the energy setting 2 (38–652 eV) and the Maxwell–Boltzmann distribution convolved with relatively wide energy resolution and response functions, respectively [from Futaana et al. (2012)]

to the deflected flow of the solar wind plasma around the mini-magnetosphere (Fig. 4).

Vorburger et al. (2012) have investigated the shielding efficiency of several magnetic anomaly regions for varying solar wind conditions and found that the shielding efficiency depends not only on the solar wind dynamic pressure but also on the geometry (structure) of the anomalies (Fig. 5). 3D kinetic particle-in-cell simulations of the solar wind with lunar crustal magnetic anomalies showed that the anomalies that are strong enough can stand-off solar wind and form mini-magnetosphere (Kallio et al. 2012; Jarvinen et al. 2014; Deca et al. 2014). The simulation showed that the interaction is mostly electron driven and that the mini-magnetosphere is unstable over time, which leads to only temporal shielding of the surface underneath. Recent hybrid simulations of solar wind interaction with magnetic anomalies (Giacalone and Hood 2015) have also shown that magnetic anomalies deflect the solar wind and that the solar wind proton flux reduces by a factor of 4 at the centre of the anomaly regions. Further, the hybrid simulation showed that the energy of the solar wind protons which hit the lunar surface within the magnetic anomaly region can be much lower than that of solar wind (10–60 % of solar wind energy) which indicates deceleration of the solar wind protons before reaching the lunar surface.

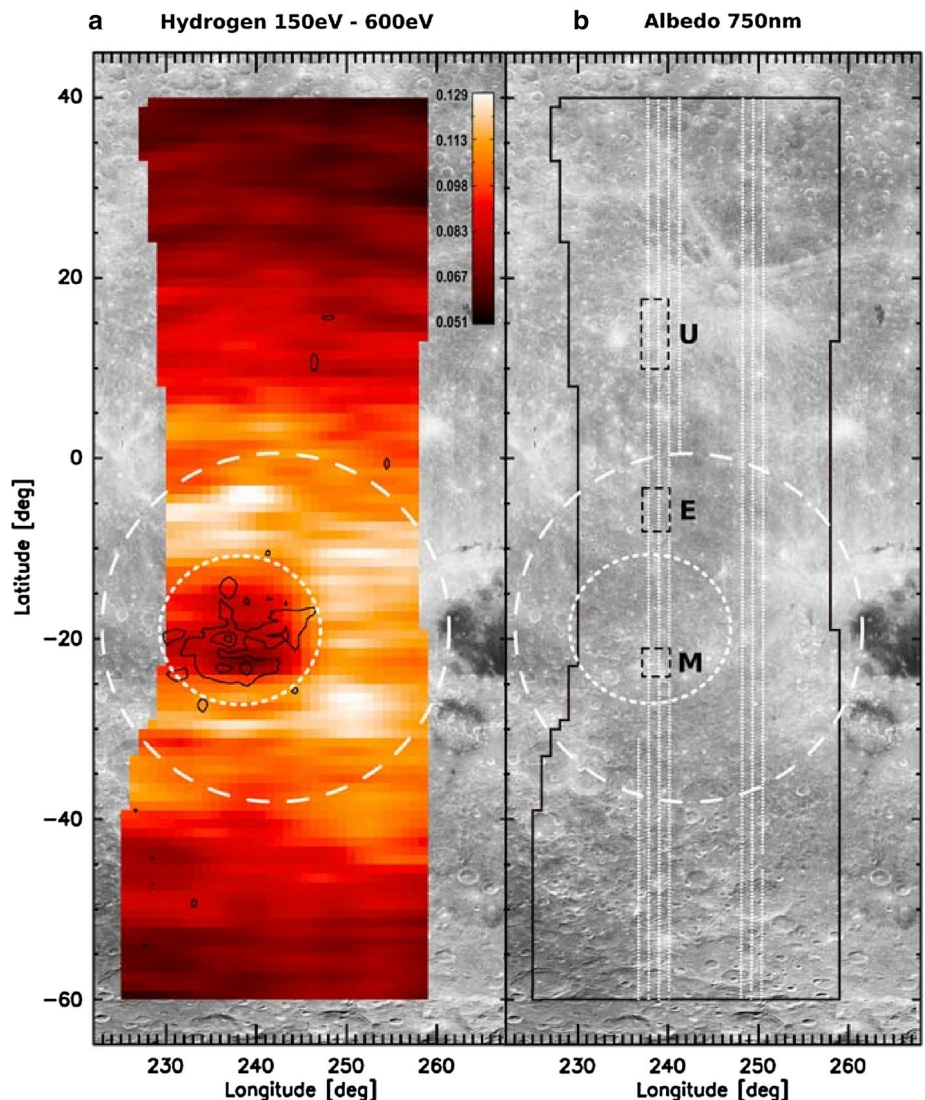

**Fig. 4** The back-scattered hydrogen ENA reflection ratio over the magnetic anomaly near 22° S and 240° E on the lunar farside observed from 200 km altitude on 17 June 2009. The maps show a unit-less reflection coefficient: neutral hydrogen number flux integrated over the specified energy range divided by total solar wind number flux integrated over energy and cosine of lunar latitude in the energy ranges **a** 150–600 eV. *Black contours* in the *centre* show the magnetic field magnitude at 30 km altitude obtained from Lunar Prospector data, with lines for 5, 15, and 25 nT. The *dotted circle* represents the region of magnetic anomaly and the *dashed circle* represents the region just surrounding the anomaly. **b** Context image taken from the Clementine *grey scale* albedo map where the regions *M*, *E*, and *U* indicate three sample regions inside the mini-magnetosphere, the enhanced flux region, and the undisturbed region, respectively [from Wieser et al. (2010)]

Futaana et al. (2013) estimated the lunar surface potential using an empirical method that uses the observed ENA intensity over a magnetic anomaly. The empirical method is based on the energy change of the impinging

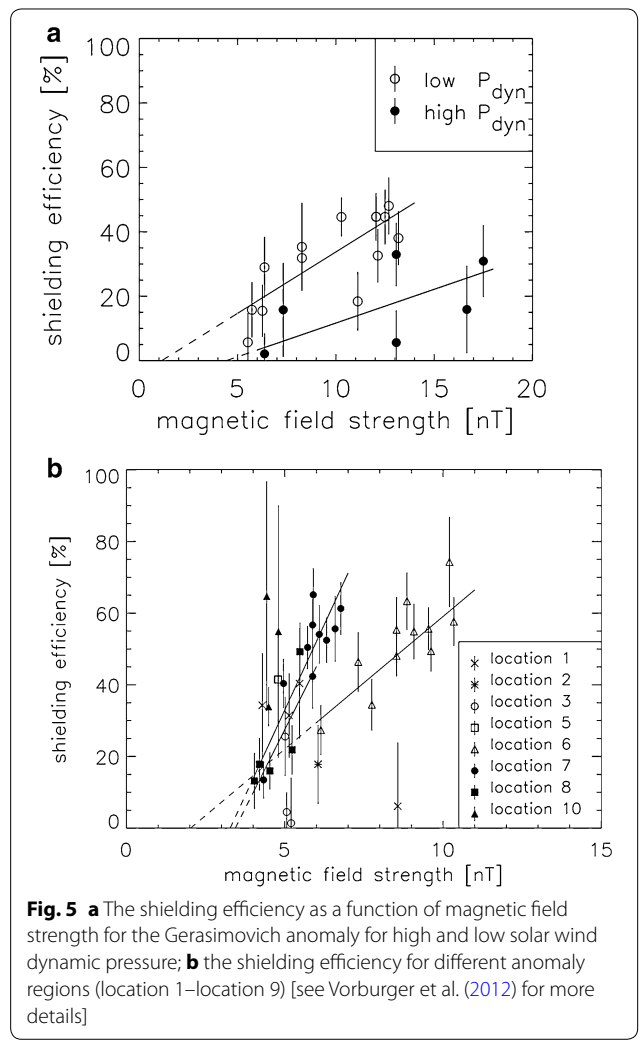

**Fig. 5** **a** The shielding efficiency as a function of magnetic field strength for the Gerasimovich anomaly for high and low solar wind dynamic pressure; **b** the shielding efficiency for different anomaly regions (location 1–location 9) [see Vorburger et al. (2012) for more details]

experiments with a flowing plasma of energy up to 55 eV interacting with a magnetic dipole field above an insulating surface, Howes et al. (2015) investigated the lunar surface charging in the magnetic anomaly regions. They have found that the surface gets charged to larger positive potentials of around +50 V when the dipole moment was perpendicular to the surface. When the dipole moment becomes parallel to the surface, the potentials were found to be much lower (around +7V).

Using the entire data of ENA observations by Chandrayaan-1/CENA, the first global ENA albedo map covering ~89 % of the lunar surface has been generated (Vorburger et al. 2013) (Fig. 7). Local variations in the ENA albedo were seen from the map, which indicated that lunar surface is not a homogeneous ENA reflector but exhibits local variations and the signatures of the anomaly regions could be clearly seen in the map. All these are indicative that the ENA imaging is a powerful tool to investigate the planetary surface.

The analysis of ENAs from the South pole Aitken basin has been carried out to investigate the effect of the surface properties on the variation of backscattered ENA albedo (Vorburger et al. 2015). South pole Aitken terrain has highly variable surface properties and magnetic anomalies around the rim. The analysis showed that the ENA albedo are sensitive to magnetic anomalies which causes the suppression of ENA intensity, whereas the other surface properties, e.g., changes in elevation of surface, chemical composition, and visible albedo do not play a significant role in the variation of ENA albedo.

Hydrogen ENAs are also found to be produced when the Moon is in the geomagnetic tail, where the plasma conditions are different compared to upstream solar wind (Allegrini et al. 2013; Harada et al. 2014). On an average, the ENA intensities are found to be higher when the Moon is in the Earth's magnetosheath (Allegrini et al. 2013) compared to that in the upstream solar wind. The energy spectra of these ENAs are found to be described well by a power law, which matches with that of ENAs observed in upstream for energies above 0.6 times the solar wind energy, but below this energy there are large differences up to a factor of 10. The velocity distribution of plasma in the magnetosheath is broader compared to that of upstream solar wind (hence broader angular distribution) and hence has smaller Mach number. This means that the area of the lunar regolith seen by the magnetosheath plasma will be higher compared to that of solar wind and hence more back-scattered ENAs. This was considered as potential factor that contributes towards the higher ENA emission in the magnetosheath region. ENAs have also been observed when the Moon was in the plasmasheet region with a backscattering fraction of ~10 % (Harada et al. 2014). The backscattering solar wind protons being decelerated by the surface potential (solar wind energy at the surface) and relies on the ENA energy spectrum and the ENA characteristic energy. The surface potential map (Fig. 6) shows larger positive potential inside the anomaly region, which is expected due to the outward electric field resulting from the higher penetration of protons compared to that of electrons in this region (Kallio et al. 2012; Jarvinen et al. 2014). However, the lack of electrostatic potential in the enhanced region [the region where the ENA intensity have been found be enhanced when the mini-magnetosphere was discovered (Wieser et al. 2010) as discussed in the paragraph above] indicates that there is no electric potential formed above the enhanced region and hence the deflection above the anomaly is mainly caused by magnetic forces. This suggests that a mini-scale bow shock, as hypothesised earlier, is evidently not formed above the magnetic anomaly. Using laboratory

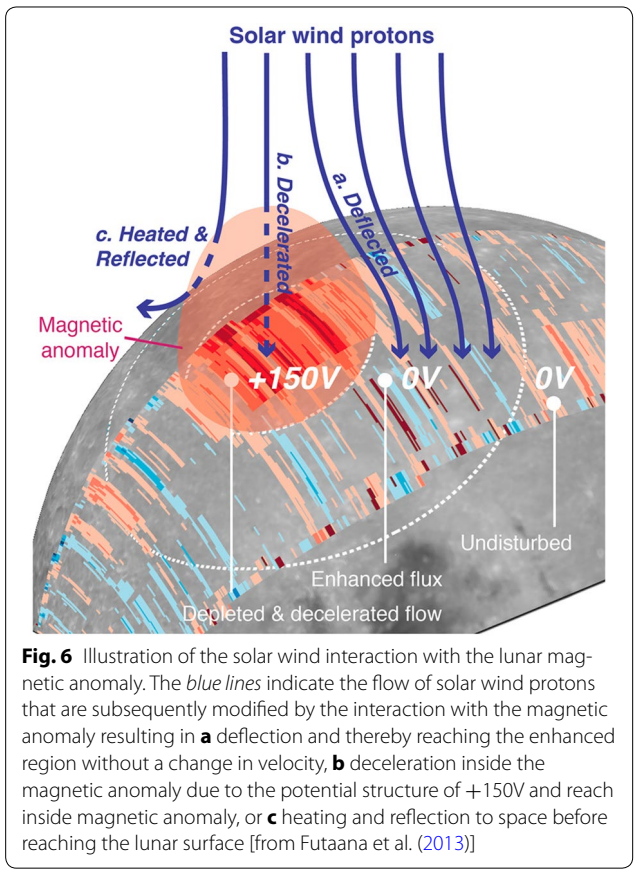

**Fig. 6** Illustration of the solar wind interaction with the lunar magnetic anomaly. The *blue lines* indicate the flow of solar wind protons that are subsequently modified by the interaction with the magnetic anomaly resulting in **a** deflection and thereby reaching the enhanced region without a change in velocity, **b** deceleration inside the magnetic anomaly due to the potential structure of +150V and reach inside magnetic anomaly, or **c** heating and reflection to space before reaching the lunar surface [from Futaana et al. (2013)]

strong magnetic anomaly region when the Moon is in the upstream solar wind. Test particle trajectory simulations carried out to investigate the effect of simple dipole magnetic fields for the backscattering process showed that the shielding of the surface by magnetic fields is less efficient for an incident beam of hot protons similar to that in the plasmasheet (Harada et al. 2014).

Solar wind has been considered as a potential contributor towards lunar exosphere by sputtering of the lunar surface, a process which has been theoretically modelled (Wurz et al. 2007 and references therein). In the recent past, Killen et al. (2012) have estimated the sputter yields for different solar wind conditions such as fast and slow solar wind, solar energetic particle population, and coronal mass ejections (CME) using Monte Carlo model. They found that the enhancement of heavier ions such as $He^{++}$ and $O^{7+}$ increases the sputter yield. Under CME passage, the mass of elements such as Na, K, Ca, and Mg in the lunar exosphere was found to increase to more than ten times their background values.

The first direct observational evidence for the solar wind sputtering of lunar surface came from CENA/SARA sensor of Chandrayaan-1 with the observation of sputtered oxygen atoms (Vorburger et al. 2014). The sputtered oxygen flux is around 0.2–0.4 times that of the backscattered hydrogen ENA flux. As expected, the sputtered flux was found to be higher when the solar wind had higher $He^{++}$ content, since $He^{++}$ is a more efficient sputtering agent compared to $H^+$ due to its higher mass (Fig. 8). The density of oxygen atoms at subsolar point was estimated to be $(1.1 \pm 0.3) \times 10^7$ m$^{-3}$ for solar wind with low Helium content (<0.35 %) and $(1.4 \pm 0.4) \times 10^7$ m$^{-3}$ for high Helium content (~4.57 %), which agrees well with fraction in the plasmasheet does not show significant difference between northern and southern hemispheres, whereas back-scattering fraction was found to reduce to almost 50 % in southern hemisphere over a large and model predictions and observational upper bound. Weak signals of Helium ENAs have also been detected for the first time (Vorburger et al. 2014).

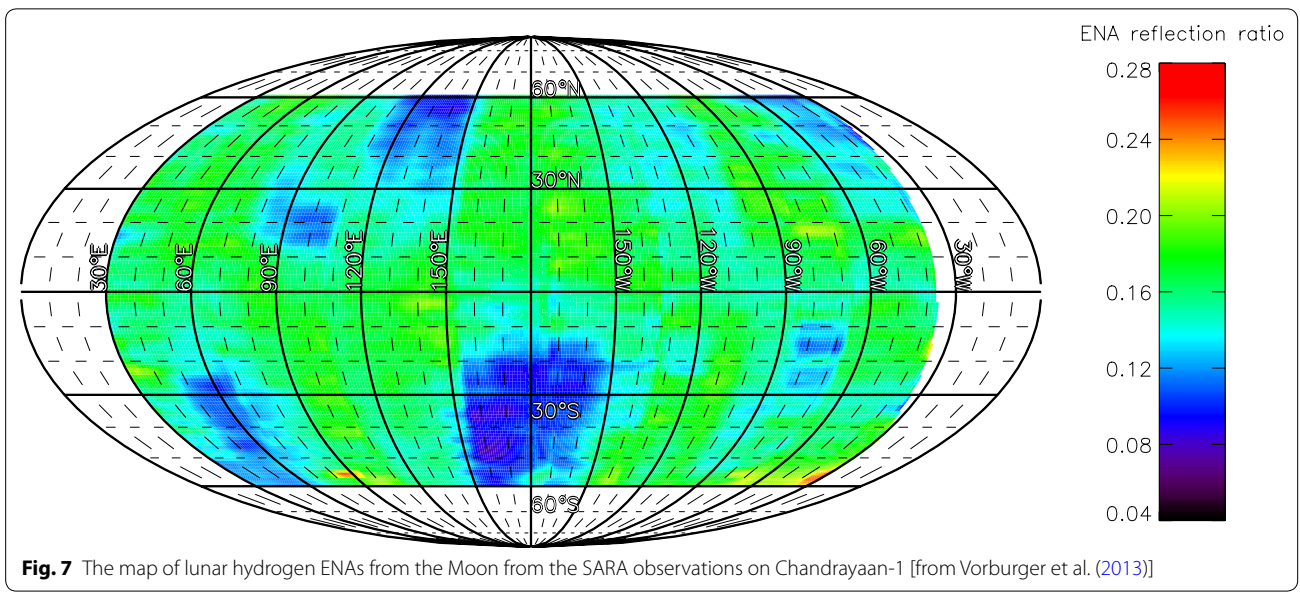

**Fig. 7** The map of lunar hydrogen ENAs from the Moon from the SARA observations on Chandrayaan-1 [from Vorburger et al. (2013)]

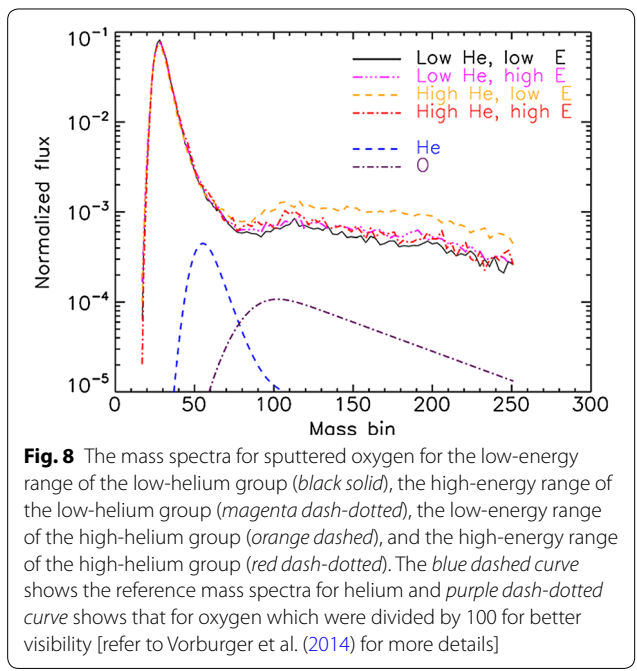

**Fig. 8** The mass spectra for sputtered oxygen for the low-energy range of the low-helium group (*black solid*), the high-energy range of the low-helium group (*magenta dash-dotted*), the low-energy range of the high-helium group (*orange dashed*), and the high-energy range of the high-helium group (*red dash-dotted*). The *blue dashed curve* shows the reference mass spectra for helium and *purple dash-dotted curve* shows that for oxygen which were divided by 100 for better visibility [refer to Vorburger et al. (2014) for more details]

There have been several efforts towards modelling of the lunar exosphere which were later compared with measurements made by LRO and LADEE. Sarantos et al. (2012), also modelled the column abundances of expected lunar exospheric constituents and found that except for Ca, the modelled abundances matches with the available measurements. This suggested the role of loss processes such as gas-to-solid phase condensation during micrometeoroid impacts or the formation of stable metallic oxides, which were not included in the model. In addition, Sarantos et al. (2012) have compared the estimates of the ion flux of the expected lunar exospheric constituents with the estimates for the ion fluxes ejected from the lunar surface by solar wind ions and electrons. This showed that measurements of ions around Moon will help to understand the abundances of many undetected species in the lunar exosphere because the expected ion flux levels from the exosphere exceed those from the surface. The observations of He in the lunar exosphere using the Lyman Alpha Mapping Project (LAMP) ultraviolet spectrograph on LRO showed day-to-day variations in the derived surface He density which correspond to the variations in the solar wind $He^{++}$ flux (Feldman et al. 2012). In addition, there was a reduction by a factor of two in the surface He density when the Moon passes through Earth's magnetotail, which was explained by models.

Recently, Ne was discovered in the lunar exosphere from the Neutral Mass Spectrometer (NMS) onboard the LADEE, and the first global characterization of He and Ar was also obtained from the observations (Benna et al. 2015). They found that He is controlled by the solar wind alpha particles and also by a endogenous source that supplies at a rate of $1.9 \times 10^{23}$ atoms s$^{-1}$. Neon was detected over the night side and the levels were comparable to that of He. Ne was found to exhibit the spatial distribution of a surface-accommodated non-condensible gas. Localized Ar enhancement was also observed at the western maria which was never observed before and the variability resulting from this local enhancement would couple to a more global source that is transient in nature. Using the data from the CHACE experiment on Chandrayaan-1 mission, Thampi et al. (2015) have derived the 2-D distribution of the lunar molecular hydrogen ($H_2$). By comparing with the topography data from the Lunar Laser Ranging Instrument (LLRI) on Chandrayaan-1, they concluded that the surface processes play an important role in the small-scale variations of the $H_2$ number density in tenuous lunar atmosphere. Based on the observations of Lunar-based Ultraviolet Telescope on Chang'E-3, an upperlimit of $10^{11}$ cm$^{-2}$ for the column density of OH radical and $10^4$ cm$^{-3}$ for the OH surface concentration has been reported (Wang et al. 2015). This estimation was based on the assumption of a hydrostatic equilibrium with a scale height of 100 km and that the recorded background to be fully contributed by the resonance fluorescence emission. The concentration values were found to be lower than the previous reports by about two orders of magnitude, whereas it was closer to those predicted by the sputtering models.

Using soft X-ray imaging of the Moon from the Röntgen satellite (ROSAT), Collier et al. (2014) found that the increase in the soft X-ray intensity associated with the limb brightening is consistent with that expected from the solar wind charge exchange with the lunar exosphere based on the lunar exospheric models and hybrid simulation results of solar wind access beyond the terminator. Since the signal appears to be dominated by exospheric species arising from solar wind implantation, the soft X-ray imaging can be used to infer the lunar limb column density as well as to determine how the exosphere varies with solar wind conditions. In addition, using the ground-based measurements from the National Solar Observatory McMath-Pierce Telescope, Mierkiewicz et al. (2014) have investigated the temperature and velocity variation of the lunar sodium exosphere. They found effective temperatures ranging between $3260 \pm 190$ K closer to sunlit east and $1000 \pm 135$ K closer to south lunar limbs and velocity in the range 100–600 m s$^{-1}$ between different locations off the lunar limb.

**Plasma environment around the Moon**

The electron and magnetic field measurements by Lunar Prospector (Lin et al. 1998; Halekas et al. 2001, 2005; Hood et al. 2001; Richmond and Hood 2008; Tsunakawa et al. 2015 and references therein) provided a map of the magnetic field over the lunar surface which clearly showed the magnetic anomaly regions and extensive investigations about the electrons in the near-lunar wake region, as well as the lunar surface charging. In the recent years, the ion measurements around Moon have been made by the plasma analysers onboard Kaguya, Chang'E-1, Chandrayaan-1, and ARTEMIS.

Mainly three ion populations have been observed as a result of the interaction of the solar wind with the day-side lunar surface: (1) solar wind protons scattered from the lunar surface, (2) protons scattered from lunar magnetic anomalies, and (3) ions of lunar origin. About 0.1–1 % of the solar wind protons are found to scatter back preserving the charge state after interactions with the regolith (Saito et al. 2008; Holmström et al. 2010). These scattered protons can be further accelerated by the convective electric field (Futaana et al. 2010; Saito et al. 2010) and hybrid simulations showed that they contribute to the plasma environment around Moon (Holmström et al. 2010). The study of the surface-scattered protons using the Chandrayaan-1/SWIM showed that the backscatter fraction is in the range 0.1–1 % and that the backscattering efficiency depends largely on the solar wind velocity indicating that the scattering is momentum-driven process (Lue et al. 2014) (Fig. 9).

Apart from scattering from surface, protons deflected from the magnetic anomaly region have also been observed (Saito et al. 2010; Lue et al. 2011). The deflection efficiency was found to be around 10 % on average whereas at the location of strongest anomalies, it could be as high as 50 %. The energy of these deflected protons is about the same as that of solar wind, but the spectrum is broader indicating some heating effects. The heating has also been observed for electrons which are observed simultaneously during the proton reflection (Saito et al. 2010). The investigation of accelerated ions in the 0.23–1.5 keV range around the magnetic anomaly region in the South-pole Aitken basin observed by Kaguya (Yokota et al. 2014) suggested that the electric field produced due to the interaction of solar wind with the magnetic anomaly is the source of the acceleration. The ions included the ones that are scattered from the surface as well as those of solar wind origin. However, structures similar to those associated with collisionless shocks have been observed near the Moon recently by ARTEMIS (Halekas et al. 2014). The location of these structures with regard to the magnetic anomaly regions suggests that the solar wind protons reflected from the anomaly may be the source of the observed small-scale shock. Using a 3-D hybrid simulation, Jarvinen et al. (2014) showed the presence of a potential of <300 V at the lunar surface which is associated with the Hall electric field in the anti-Moonward direction that results from the interaction between solar wind and a magnetic dipole mimicking the Gerasimovich magnetic anomaly. This value is consistent with the value of >135 V reported by Futaana et al. (2013) using ENAs at Gerasimovich magnetic anomaly and ~150 V reported by Saito et al. (2012) from the ion and electron observation in the south-pole Aitken region at 25 km altitude.

There have been observations of heavier ions originating from the lunar surface/exosphere (Yokota et al. 2009). The ions observed are $He^+$, $C^+$, $O^+$, $Na^+$, and $K^+$. These are indicative of the presence of He, O, and C exospheres around the Moon. The longitude distributions of $Na^+$ and $K^+$ fluxes from the Kaguya low-energy ion data, showed a dawn–dusk asymmetry such that the abundance decreases to approximately 50 % at dusk compared with that at dawn (Yokota et al. 2014). This can be due to the emission of the exospheric particles at dawn side which also implied that the surface abundance of Na and K need to be supplied during the night to explain the observed dawn–dusk asymmetry. The pickup of the exospheric particle with $M/q = 2$ has also been observed with flux of $\sim 10^{-5}$ to $10^{-4}$ of the solar wind flux. $H_2^+$ is considered as the most probable candidate for the observed species (Wang et al. 2011). The flux of the observed pickup ions are in the range $10^{-5}$–$10^{-4}$ of the solar wind flux. The energy of the pickup ions has been found to depend not only on the location of detection, but also on the incident angles from the observations of Chang'E-1 (Zhong et al. 2013). In addition, these authors found the acceleration of the ions to be less efficient when the component of interplanetary magnetic field perpendicular to the solar wind flow in the ecliptic plane (IMF $B_y$ in Geo-centric Solar Ecliptic frame) was reduced. This shows the control of IMF in the acceleration mechanism. Pickup ions have been observed near the Moon when the Moon was in the magnetotail lobes of Earth (Poppe et al. 2012) as well as upstream (Halekas et al. 2012). The ions observed in the terrestrial lobe are inferred to originate from lunar exospheric particles that are ionised in the tail-lobe. The mass of the ions observed in the upstream is in the range 20–45 amu and the pickup ion flux correlated with the solar wind flux suggesting sputtering as a key mechanism in producing the observed ions either as ions directly from the surface or as neutrals that are subsequently ionised (Halekas et al. 2012). From the observation of pick-up ions by ARTEMIS on the dawn side of the terrestrial magnetosheath, the abundance of the species of mass 16 amu was found to be much higher than expected

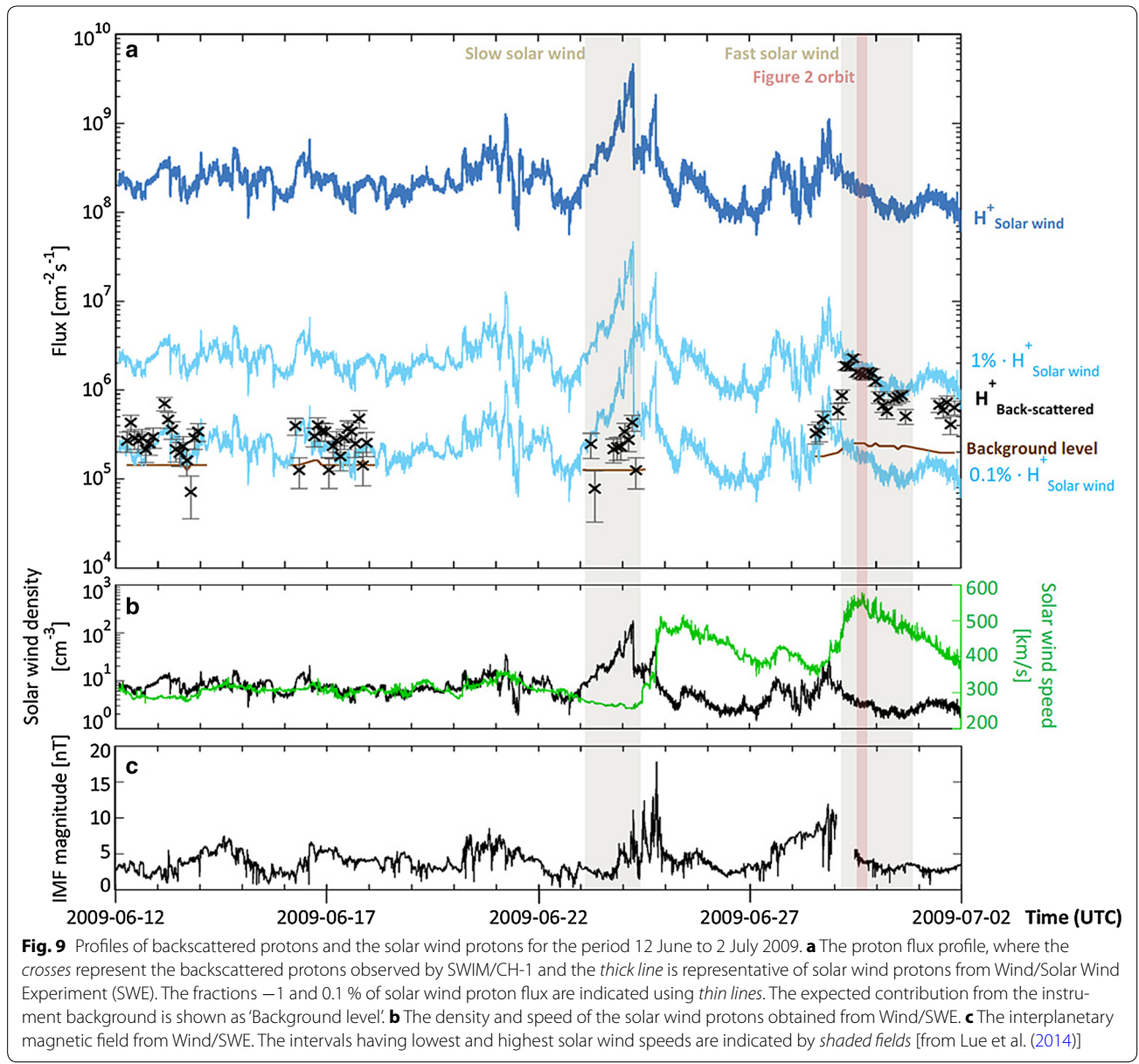

**Fig. 9** Profiles of backscattered protons and the solar wind protons for the period 12 June to 2 July 2009. **a** The proton flux profile, where the *crosses* represent the backscattered protons observed by SWIM/CH-1 and the *thick line* is representative of solar wind protons from Wind/Solar Wind Experiment (SWE). The fractions −1 and 0.1 % of solar wind proton flux are indicated using *thin lines*. The expected contribution from the instrument background is shown as 'Background level'. **b** The density and speed of the solar wind protons obtained from Wind/SWE. **c** The interplanetary magnetic field from Wind/SWE. The intervals having lowest and highest solar wind speeds are indicated by *shaded fields* [from Lue et al. (2014)]

(Halekas et al. 2013). This could have been due to contributions from terrestrial oxygen, OH, or CH$_4$.

Regarding the nightside plasma population, recent observations from plasma analysers onboard Kaguya, Chang'E-1, Chandrayaan-1, and ARTEMIS, have shown the presence of protons in the near-lunar wake region (100–200 km above the surface) (Halekas et al. 2015). This region was thought to be devoid of any plasma population. Instead, protons were found to access the near-lunar wake by travelling parallel as well as perpendicular to the interplanetary magnetic field (IMF). The sources for the perpendicular entry are identified as (1) surface-scattered solar wind protons near the dayside equator, accessing the near-lunar wake under the influence of the convective electric field (Nishino et al. 2009), (2) the forward-scattered protons near the terminator (Wang et al. 2010), the increase in the Larmor radius of protons under the influence of wake boundary electric field thereby accessing the near wake (Nishino et al. 2009), and (3) the protons from the tail of the solar wind velocity distribution that have gyro-radii larger than the radius of the Moon and can directly enter the lunar wake region (Dhanya et al. 2013) (Fig. 10). The parallel entry of the protons was observed by Chandrayaan-1 (Futaana et al. 2010) at a distance of 100 km from the lunar surface and by ARTEMIS at a distance of around 3.5 $R_L$

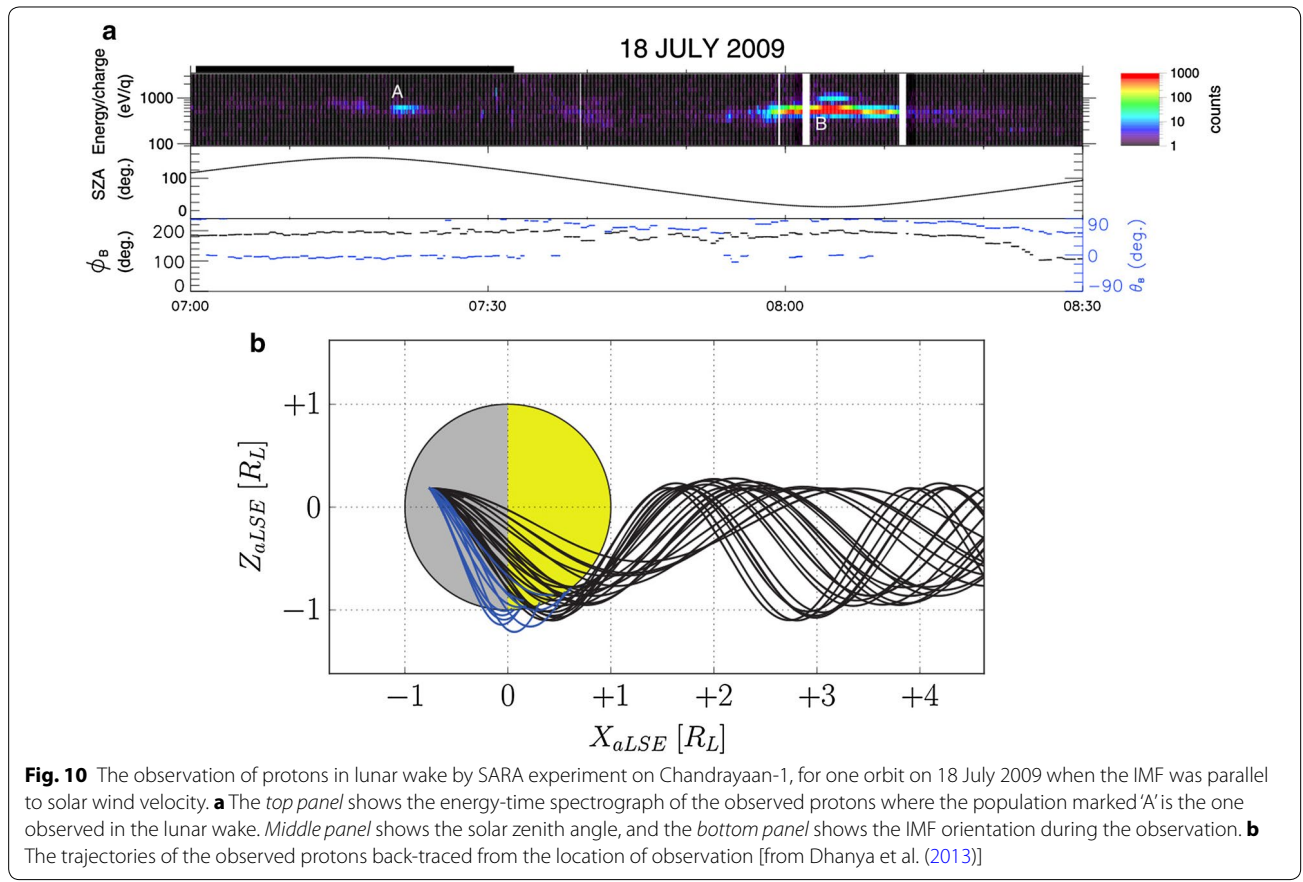

**Fig. 10** The observation of protons in lunar wake by SARA experiment on Chandrayaan-1, for one orbit on 18 July 2009 when the IMF was parallel to solar wind velocity. **a** The *top panel* shows the energy-time spectrograph of the observed protons where the population marked 'A' is the one observed in the lunar wake. *Middle panel* shows the solar zenith angle, and the *bottom panel* shows the IMF orientation during the observation. **b** The trajectories of the observed protons back-traced from the location of observation [from Dhanya et al. (2013)]

from the Moon (Wiehle et al. 2011), where $R_L$ is the lunar radius ($R_L \sim 1738$ km). The 1-D plasma expansion model (Gurevich et al. 1966; Samir et al. 1983) was able to explain the ARTEMIS observations, whereas it is not consistent with the Chandrayaan-1 observations.

In addition to protons, $He^{++}$ have also been observed in lunar wake (Halekas et al. 2011). Furthermore, the electron flux enhancement has been reported in the lunar wake above the Crisium antipode magnetic anomaly region (Nishino et al. 2015) and are characterised by low energy bi-directional field-aligned electron beams and a medium-energy population which are linked to the bouncing motion between the footprints of the crustal field. The theory behind the lunar wake refilling along the magnetic field (one-dimensional solution) has been discussed for background electron distributions ranging from Maxwellian to Kappa distributions (Halekas et al. 2014). Recent observations by ARTEMIS have shown magnetic field enhancements in the wake to more than twice that of ambient IMF during high plasma-beta conditions (Poppe et al. 2014b).

Recently, Kaguya observations on the lunar nightside when the Moon is in Earth's plasmasheet have shown that the plasmasheet ions are accelerated by the spacecraft potential which is negatively charged (the electrons are accelerated by the potential difference between the spacecraft and the lunar nightside surface) (Saito et al. 2014). Using this, the nightside surface potential was estimated in the range −100 to −500 V.

## Discussion and conclusions

The latest understanding on the various processes initiated by the interaction of solar wind with the Moon is summarized in Fig. 11. Such processes include, but not limited to, backscattering of solar wind protons as hydrogen ENAs, sputtering of the lunar surface, scattering of solar wind protons from the lunar surface as well as from magnetic anomalies, nightside ions, plasma dynamics in near-lunar wake.

The ENA backscattering as well as proton scattering from lunar surface indicates the presence of complex microphysics behind the plasma interaction with the regolith. The observation of neutralisation of solar wind protons with a porous regolith and backscattering with energies closer to that of solar wind calls for a need to understand the microphysics behind the processes that must have occurred on the lunar surface since its formation. The larger sunward scattering of ENAs for larger solar zenith angle suggests that the underlying processes are different compared to that observed in the laboratory (Niehus et al. 1993). The ENA emission is not only a global phenomena on Moon, but also occurs under different plasma regimes, such as the magnetosheath and the plasmasheet.

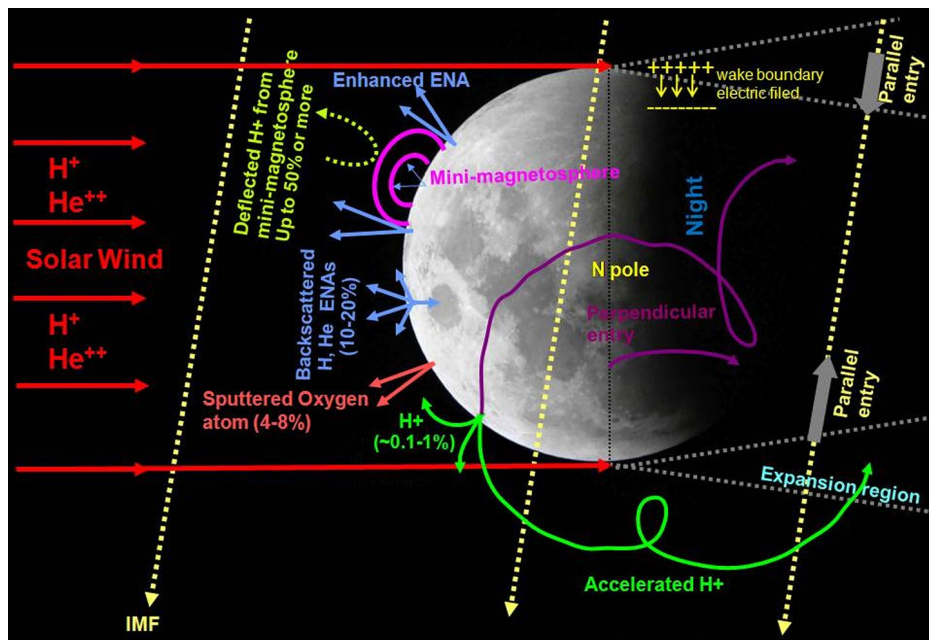

**Fig. 11** A summary of our current understanding of the processes initiated by the interaction of solar wind with the Moon

The surface of the Moon undergoes space weathering due to the continuous bombardment by high-energy solar wind plasma (Denevi et al. 2014 and references therein). The shielding of the surface from solar wind plasma by magnetic anomalies shows that the weathering is not uniform across the surface. The regions inside anomalies undergo less weathering and the region surrounding the local magnetic field received more solar wind flux. Thus the maturity of the soil is different at different locations on the Moon. This variable weathering is seen in the spectral features of the regions around the magnetic anomalies such as the high albedo regions known as 'swirls' (Blewett et al. 2011 and references therein). It has been found that most of the observed swirls have been associated with magnetic anomalies and are located inside the anomalies such as swirls of Reiner Gamma and the swirls in Mare Ingenii and Mare Marginis. The compositional analysis of several swirl regions by Blewett et al. (2011) showed the spectral characteristics of swirls to be similar to that of immature material and swirls were found to have slightly lower FeO values compared with their surroundings. The solar wind shielding by the magnetic anomalies were identified as one of the potential sources of this compositional difference.

The spectral analysis by the Moon Mineralogy Mapper of Chandrayaan-1 has found that the maturity of the swirl region is low compared to surrounding regions (Kramer et al. 2011) and also that the swirls are depleted in OH relative to their surrounding regions. The UV spectra of lunar swirls were found to be that of immature, crystalline material as shown by the investigation by LROC (Denevi et al. 2014). Recently, the LRO data (Timothy et al. 2015) have supported the hypothesis that swirls are formed as a result of deflection of the solar wind by local magnetic fields. The comparison of high-albedo regions on Moon with the latest vector magnetic field map of the magnetic anomalies from Kaguya and Lunar Prospector observations (Tsunakawa et al. 2015) have suggested that 3-D structure of near-surface magnetic fields is important for the swirl formation at the magnetic anomaly. Thus, imaging of the lunar surface using ENAs has proved to be a powerful technique to detect local magnetic fields (magnetic anomaly regions) and mini-magnetospheres. This adds to the understanding of surface weathering.

Based on observations (Pieters et al. 2009; Sunshine et al. 2009; McCord et al. 2011) as well as laboratory simulations (Managadze et al. 2011; Izawa et al. 2014; Schaible and Baragiola 2014), the implantation of solar wind protons has been considered to be a potential source for the formation of OH on lunar surface. Recent modelling studies of the implantation (Farrell et al. 2015) of solar wind protons into the surface using a Monte Carlo approach showed that implanted $H^+$ is retained on the surfaces with activation energy >1 eV and the retained $H^+$ are likely to form OH. In recent laboratory simulations by Schaible et al. (2014), amorphous $SiO_2$ and olivine samples were bombarded by proton beams of 2–10 keV under ultrahigh vacuum condition and it was found that about 90 % of the implanted $H^+$ gets converted to hydroxyl species initially and the rate of conversion decreased exponentially with fluence. From this, it was estimated that on the lunar regolith, the solar wind-implanted $H^+$ can account for the 17 % of the relative OH absorption which is consistent with observations. The backscattering of solar wind by ∼20 % reduces the fraction of hydrogen available for OH formation and may affect the amount of OH. In addition, the shielding of the lunar surface by magnetic anomalies shows that the solar wind implantation will be less at regions protected by anomalies and hence, a more pronounced depletion of the OH content in the anomaly region (Kramer et al. 2011). Recent investigations using Monte Carlo modelling have shown that the magnetic anomalies suppress the sputtered intensity thereby causing anisotropies in the lunar exosphere (Poppe et al. 2014a).

The different charged particle populations such as the surface-scattered solar wind protons on the dayside that are subsequently accelerated by the solar wind convective electric field, the suspected exospheric ions, and the protons with different velocity distributions co-existing in the lunar wake, affect the lunar plasma environment. The protons on the nightside may alter the surface charging and even cause sputtering of the nightside lunar surface. The 3-D hybrid simulations (Fatemi et al. 2013) have shown the existence of three current systems in lunar wake in the transition regions between the plasma void, the rarefaction region, and the interplanetary plasma. These currents can induce the magnetic field and perturb the existing field lines and these currents were found to depend on the direction of the interplanetary magnetic field direction. Poppe et al. (2013) have modelled the possibility of a sputtering process initiated by the lunar exospheric ions, a process termed as self-sputtering. The model calculations suggested that the rate at which this processes sputter the particles may be equal or exceed that due to solar wind sputtering and micrometeoroid bombardment. Also, the rate is higher than that of sputtering due to passage through the Earth's plasma sheet.

Significant wave activities also have been initiated due to the different plasma population around Moon. A proton-governed region is formed on the nightside that subsequently creates an outward electric field, which accelerates the electrons inward (Nishino et al. 2010). The abundance of positive charge in the range $10^{-5}$ to $10^{-7}$ $cm^{-3}$ can accelerate electrons up to 1 keV and broadband electrostatic noise due to the counter-streaming electron beams has been found as a result of this process. Electrostatic solitary waves and whistler waves have been observed around Moon (Halekas et al. 2006; Tsugawa et al. 2011, 2012, 2015; Nakagawa et al. 2011). Electrostatic solitary waves has been observed on the day-side lunar surface as well as at the boundary of the lunar wake from Kaguya (Hashimoto et al. 2010). At the wake boundary the waves were found to be generated due to the electrons which are reflected and accelerated by an electric field at the boundary. On the day-side strong wave activities were produced by electrons which are mirror reflected at the magnetic anomalies and also weaker component due to electrons that are scattered from the lunar surface.

Whistler waves have been observed around Moon from Lunar Prospector in the frequency range 0.4–4 Hz (Halekas et al. 2006 and from Kaguya in the frequency range 0.1–10 Hz (Nakagawa et al. 2011; Tsugawa et al. 2011, 2012). From Lunar Prospector, the whistler waves were observed on the dayside, outside the lunar wake, and suggested a possible link of solar wind interaction with lunar magnetic anomalies as the source (Halekas et al. 2006). The analysis of the narrow-band whistler waves observed as magnetic fluctuation close to 1 Hz from Kaguya (Tsugawa et al. 2011) suggested the ion beams reflected by the lunar magnetic anomalies as a possible source of the waves. Nakagawa et al. (2011) reported the non-monochromatic fluctuations of the magnetic field in the frequency range of 0.03–10 Hz around Moon from Kaguya. These fluctuations, which were supposed to be the whistler waves, were observed on the lunar dayside as well as near the terminator region and were not really linked to magnetic anomalies. The source of these waves was considered as the solar wind protons that are scattered from lunar surface with wider range of velocity component parallel to the magnetic field. From Kaguya observations, Tsugawa et al. (2012) have reported the broadband whistler waves that are essentially produced by the interaction of solar wind lunar magnetic anomalies. The broadband waves and narrow band waves were observed in the same region. Their analysis suggested that broadband and narrow band waves are different views of the same waves propagating in the solar wind frame. In addition, the harmonics of the whistler waves have also been observed around lunar magnetic anomalies near the terminator region of Moon from Kaguya (Tsugawa et al. 2015).

Ultra low frequency (ULF) monochromatic waves in the dominant frequency range $8.3 \times 10^{-3}$ to $1.0 \times 10^{-2}$ Hz with large amplitudes have also been detected around the lunar dayside as well as terminator region from Kaguya when Moon was in the upstream solar wind flow (Nakagawa et al. 2012). The possible source mechanisms of these waves were considered as the cyclotron resonance between the magnetohydrodynamic waves and the solar wind protons that are reflected from the Moon. The energy of the reflected protons can account for the energy of the ULF waves. Narrow band (0.04–0.17 Hz) ion-cyclotron waves have been reported to be present at the lunar surface when the Moon was in the terrestrial magnetotail, based on the observation of Apollo 15 and 16 Lunar Surface Magnetometers. The observed differences in wave amplitude and phase at the Apollo 15 and 16 sites suggested that the wave signals were modified by the mini-magnetosphere above the Apollo 16 site. All these observations show that the electromagnetic environment around the Moon is dynamic.

The recent new findings have improved our understanding of the evolution of the Moon which has been exposed to the space environment (solar wind plasma, solar radiation, meteors, cosmic rays etc.) over billions of years since its formation. This applies to any Moon-like planetary body in our solar system as well as to exoplanets and planetary systems in any galaxy in the universe. However, the underlying physical mechanism involved in the backscattering of larger fraction of the solar wind as ENAs remains poorly understood. The future areas of investigation would include the microphysics of solar wind–regolith interaction, which apart from throwing light on the solar wind scattering as neutrals and ions itself, will also help us to understand the role of surface properties in determining the ENA energy spectra and scattering function. Contribution of the sputtered elements other than Oxygen, e.g., Ca, Si, Al, and Mg, to the lunar exosphere is not yet known, which calls for ENA sensors with much higher sensitivity in future missions. Another aspect which need attention is the interaction of solar wind with the magnetic anomalies on Moon—the fields generated due to these at different anomalies, the role of geometry of the anomalies in the interaction process, and a comprehensive knowledge on how they globally affect the neutral and plasma environment including the lunar wake. In the lunar wake, a better understanding has to evolve on the effect of the different energetic particle population on the lunar wake dynamics, surface charging, possibility of nightside sputtering, and also on how the different proton entry mechanisms interact with each other. The knowledge on the variabilities of the neutral and plasma environment in lunar day and night, as well as under different solar wind conditions, is essential to characterise the environment around Moon and for any future human base on Moon.

**Abbreviations**
ENA: energetic neutral atom; IMF: interplanetary magnetic field.

**Author details**
[1] Space Physics Laboratory, Vikram Sarabhai Space Centre, Trivandrum 695022, India. [2] Swedish Institute of Space Physics, Kiruna, Sweden. [3] Physikalisches Institut, University of Bern, Bern, Switzerland. [4] Division of Physical Sciences, American Museum of Natural History, New York, USA. [5] Space Sciences Laboratory, University of California, Berkeley, CA, USA. [6] Institute of Space and Astronautical Science, Sagamihara, Japan.

**Acknowledgements**
The efforts at Space Physics Laboratory, Vikram Sarabhai Space Centre, were supported by the Indian Space Research Organisation (ISRO). The efforts at the Swedish Institute of Space Physics were supported in part by the European Space Agency (ESA) and the Swedish Research Links Programme funded by the Swedish International Development Cooperation Agency (SIDA). The efforts at the University of Bern were supported in part by ESA and by the Swiss National Science Foundation.

**Compliance with ethical guidelines**

**Competing interests**
The authors declare that they have no competing interests.

Received: 25 March 2015   Accepted: 3 August 2015
Published online: 22 August 2015